\documentclass[aps,prl,floatfix,showpacs,twocolumn
]{revtex4}

\usepackage{amsmath}
\usepackage{graphicx}

\newcommand{\bra}[1]{\mbox{$\langle #1|$}}
\newcommand{\ket}[1]{\mbox{$|#1\rangle$}}

\newcommand{\Ket}[2]{|#1\rangle^{(#2)}}

\begin{document}

\title{Loss Tolerant Optical Qubits}

\author{T.~C.~Ralph, A.~J.~F.~Hayes, Alexei Gilchrist}
\address{Centre for Quantum
Computer Technology, Department of Physics, University of Queensland, QLD 4072, Brisbane,
Australia. alexei@physics.uq.edu.au}



\date{\today}

\begin{abstract}
	We present a linear optics quantum computation scheme that employs a new
	encoding approach that incrementally adds qubits  and is tolerant to photon loss errors.
	The scheme employs a circuit model but uses techniques from cluster state
	computation and achieves comparable resource usage. To illustrate our
	techniques we describe a quantum memory which is fault tolerant to photon
	loss.
\end{abstract}


\maketitle

Quantum logic gates can be built using linear optics, photon detection and
ancillary resources in a scalable manner, as shown by Knill, Laflamme and
Milburn (KLM) \cite{KLM01}. A number of experimental efforts are currently
focused on testing the building blocks of linear optical quantum computing
(LOQC) \cite{PJF03,Obr03,Wal04}. However, optimism for large scale quantum
computation based on LOQC has been tempered by the major overheads inherent in
the KLM scheme and the high detector and source efficiencies apparently
required \cite{KLM01}.

An alternative approach to implementing LOQC was proposed by Nielsen
\cite{Nie04} and further developed by Browne and Rudolph \cite{Bro04}
(see also \cite{yor03} for related work). This
approach combines the model of cluster-state quantum computation \cite{briegel}
with the non-deterministic gates presented by KLM, and achieves  a very
significant reduction in the overheads.  The fault tolerance of the scheme has
also been studied \cite{0405134}.


In this paper we present a new approach to LOQC based on an incremental parity
encoding \cite{Hay04}.  Our method combines ideas from both the KLM and the
cluster-state approaches. Parity encoding was used in the original KLM proposal
to protect against both teleporter failures (i.e. the non-determinism of the
gates) and photon loss. By using parity encoding but re-encoding incrementally
(instead of by concatenation) we can obtain the reduction in overheads
characteristic of the cluster state approach whilst retaining the photon loss
tolerance of KLM. 

In particular we will describe a quantum memory which is fault tolerant
\cite{mikenike} to photon loss. Though our techniques for detecting and
correcting loss are themselves themselves also subject to loss, above a
particular threshold efficiency the effect of loss can be negated to arbitrary
accuracy.  A previous description of an optical quantum memory based on error
correction did not consider fault tolerance \cite{Gin03}.
Although we specifically only consider memory, our
construction is compatible with gate operations and thus can form a template
for fault tolerant quantum computation with respect to photon loss. 
We will deal with qubits in three different tiers of encoding:
\emph{physical} encoding, \emph{parity} encoding and \emph{redundant} encoding.
The application we describe in this letter operates at the redundant-encoding
level to protect information from photon loss.

\textbf{Physical encoding:} 
At the first tier are the basic physical states that we will use to construct
qubits, these will be the polarisation states of a photon so that $\ket{0}
\equiv\ket{H}$ and $\ket{1}\equiv\ket{V}$.   The advantage of this choice in
optics, is that we can perform any single physical-qubit unitary
\emph{deterministically} with passive linear optical elements.
Of course gates
between different physical qubits become difficult and in LOQC these are
non-deterministic.

\textbf{Parity encoding:}
at the second tier of encoding are \emph{parity qubits} encoded across many
\emph{physical qubits}.  We shall use the notation $\Ket{\psi}{n}$ to mean the
logical state $\ket{\psi}$ of a qubit, which is parity encoded across $n$
physical qubits. In this notation the \emph{physical qubits} are the first
level, and we will often drop the superscript for this level as was done above.  

Specifically, the parity encoding is given by
\begin{eqnarray}
\label{parity}
\Ket{0}{n} & \equiv & (\ket{+}^{\otimes n}+\ket{-}^{\otimes
n})/\sqrt{2}\nonumber \\  
\Ket{1}{n} & \equiv & (\ket{+}^{\otimes n}-\ket{-}^{\otimes
n})/\sqrt{2},
\end{eqnarray}
where  $\ket{\pm} = (\ket{0} \pm \ket{1})/\sqrt{2}$.  The main feature of
this encoding is that a computational basis measurement of any one of the
physical qubits will not destroy the logical state, but rather will reduce the
level of encoding by one.  

There are two operations which are easily performed on parity encoded
states, one is a rotation by an arbitrary amount around the $x$ axis of the
Bloch sphere (ie $X_\theta=\cos(\theta/2) I + i\sin(\theta/2)X$) \footnote{$X$,
$Y$, and $Z$ are the usual Pauli operators and an angle subscript denotes a
rotation about that axis, analogous to $X_\theta$ defined in the text.}, which can
be performed by applying that operation to any of the physical qubits; and the
other is a $Z$ operation, which can be performed by applying $Z$ to
\emph{all} the physical qubits (since the odd-parity states will acquire an
overall phase flip).  A key operation we will use is the partial Bell state
measurement \cite{Wei94,Brau95}.  This consists of mixing two physical qubits
on a polarising beam splitter followed by measurement in the
diagonal-antidiagonal basis.  A successful event occurs when a photon is
counted at each out put of the beamsplitter. An unsuccessful event occurs when
both photons appear at one of the outputs. When successful it projects onto the
Bell states $\ket{00}+\ket{11}$ and $\ket{00}-\ket{11}$. When unsuccessful it
projects onto the separable states $\ket{01}$ and $\ket{10}$, thus measuring
the qubits in the computational basis. This operation can be used to add $n$
physical qubits to a parity encoded state using a resource of $\Ket{0}{n+2}$.
We will refer to this as type-II fusion ($f_{II}$) following the nomenclature
of Brown and Rudolph \cite{Bro04}. We will discuss the production of the
required resource states shortly. The result of type-II fusion is
\begin{equation}
	f_{II}\Ket{\psi}{m}\Ket{0}{n+2}\rightarrow\left\{ \begin{array}{cl}
		\Ket{\psi}{m+n} & \mbox{(success)}\\
		\Ket{\psi}{m-1}\Ket{0}{n+1} & \mbox{(failure)}
	\end{array}\right.
	\label{encoding}
\end{equation}
When successful (with probability $1/2$), the length of the parity qubit is
extended by $n$. A phase flip correction may be necessary depending on the
outcome of the Bell-measurement.  If unsuccessful a physical qubit is removed
from the parity encoded state, and the resource state is left in the state
$\Ket{0}{n+1}$ (which may be recycled).  This encoding procedure is equivalent
to a gambling game where we either lose one level of encoding, or gain $n$
depending on the toss of a coin. Clearly if $n \ge 2$ this is a winning game.

The remaining gates in order to achieve a universal gate set (a $Z_{90}$ and a
\textsc{cnot} gate) can be efficiently performed on the parity encoded states
by making use of the encoder above and will be described elsewhere
\cite{parityklm}. The resource overhead for performing gates in this way is
approximately equal to the best quoted for cluster state encoding \cite{Bro04}. 

\textbf{Redundant encoding:}
The parity encoding has two purposes. Firstly the non-deterministic gates
which we will employ, fail by measuring the qubit in the computational
basis. Hence this code enables recovery from gate failures. Secondly, loss of a
photon can be considered a computational basis measurement in which we did not
find out the answer. Thus upon loss of a photon we know that the remaining
state is at worst a bit flipped version of the original. The final level of
encoding is a redundancy code which enables recovery from this
possibility of a bit flip. Thus at the highest level our logical qubits are
given by:
	\begin{equation}
	\ket{\psi}_L = \alpha \Ket{0}{n}_1 \Ket{0}{n}_2.....\Ket{0}{n}_q + \beta \Ket{1}{n}_1 \Ket{1}{n}_2.....\Ket{1}{n}_q
	\label{lqb}
	\end{equation}

We can create an ``encoder'' gate that correctly encodes a parity qubit by
simply fusing a more complicated resource state onto the parity qubit, namely
$\ket{0} \Ket{0}{n}_1 \Ket{0}{n}_2.....\Ket{0}{n}_q + \ket{1} \Ket{1}{n}_1 \Ket{1}{n}_2.....\Ket{1}{n}_q$. We attempt
type-II fusion of this resource onto the parity qubit, $\Ket{\psi}{n}$,
repeating till successful (on average twice) giving the (phase flip corrected)
result 
\begin{align}
& \alpha[\Ket{0}{n-k}\Ket{0}{n}_1.....\Ket{0}{n}_q + \Ket{1}{n-k}\Ket{1}{n}_1....\Ket{1}{n}_q] + \nonumber\\
& \beta [\Ket{1}{n-k}\Ket{0}{n}_1.....\Ket{0}{n}_q + \Ket{0}{n-k}\Ket{1}{n}_1.....\Ket{1}{n}_q]
\end{align}
where $0<k<n-1$ is the number of unsuccessful attempts made before fusion was
achieved. This state is made up of $q n$ ``new'' photons introduced by the
resource and $n-k$ of the ``old'' photons that made up the parity qubit.  By
measuring the old photons in the computational basis and making a bit flip (on
all new parity qubits if needed) we obtain the expected encoded state
(Eq.\ref{lqb}).  The previous universal set of gates can be used at this
highest level of encoding also.

\begin{figure}[htb]
\begin{center}
\includegraphics[width=8cm]{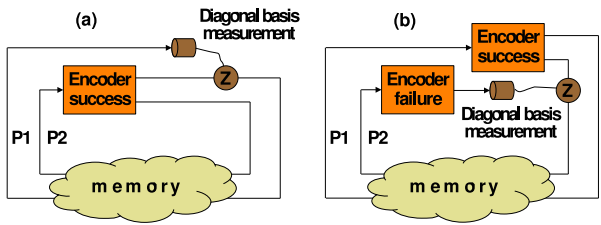}
\caption{A schematic of memory circuit. }
\label{pklm1}
\end{center}
\end{figure}

\textbf{Loss tolerant qubit memory:}
A schematic of the memory circuit for the example of a 2 qubit redundancy code
is shown in Fig.\ref{pklm1}. The basic idea is as follows. The logical qubit is
held in memory for some time as shown, during which a photon may be lost. The
logical qubit is then taken out of memory and one of its constituent parity
qubits, $P2$, is sent into the encoder described above. The encoder performs
two tasks: (i) it adds another level of redundancy encoding to the logical
qubit and; (ii) it makes a quantum non-demolition measurement of the photon
number of $P2$, which determines if a photon has been lost, without determining
the logical value of the qubit.  Fig.\ref{pklm1}(a) shows the procedure if no
photons are found to have been lost. The state straight after the encoder is:
$\ket{\psi}_L = \alpha \Ket{0}{n}_{1} \Ket{0}{n}_{2} \Ket{0}{n}_{3} + \beta
\Ket{1}{n}_{1} \Ket{1}{n}_{2} \Ket{1}{n}_{3}$ The other parity qubit, $P1$ is
now measured in the diagonal basis $\Ket{0}{n} \pm \Ket{1}{n}$. This
disentangles it from the other parity qubits which are returned to the state of
Eq.\ref{lqb} by the possible application of a phase-flip (dependent on the
outcome of the measurement on $P1$). They are returned to memory as shown.

Fig.\ref{pklm1}(b) shows the procedure when the encoder finds a photon missing
in $P2$. Now the encoded state may have suffered a bit flip and we may have the
state: $ \ket{\psi}_L = \alpha \Ket{0}{n}_1 \Ket{1}{n}_2 \Ket{1}{n}_3 + \beta
\Ket{1}{n}_1 \Ket{0}{n}_2 \Ket{0}{n}_2 \label{llqb2} $ However, recovery is
possible by now measuring the modes produced by the encoder in the diagonal
basis. This disentangles $P1$ from the other parity qubits without disturbing
its logical value. Importantly the correct $P1$ is obtained regardless of
whether a bit flip occurred to $P2$, though again a phase flip on $P1$ may be
required dependent on the outcome of the diagonal basis measurements. Finally,
$P1$ is put through an encoder, sent back to memory and the sequence is
repeated. This will correct photon loss errors in which up to a single photon
is lost per sequence. Higher levels of loss can be tolerated by increasing the
size of the redundancy code placed in memory and generalizing the protocol. For
example a 3 qubit code could be kept in memory and 3 qubit encoders used. Then
two loss events could be tolerated with recovery from the third qubit.
We will describe how the various
operations required for the memory circuit can be achieved using only linear
optics, feedforward and Bell state resources. 

\textbf{Threshold:}
Firstly consider the effect of photon loss in the encoder.
If a loss event occurs in the fusion process, that is, only one photon is
detected when a fusion is attempted, then the process is aborted. The presence
of the redundancy code allows the following recovery. One of the remaining old
photons is measured in the diagonal basis. This disentangles the entire parity
qubit on which the encoder was attempted from the other parity qubit as
described earlier. If fusion is successful but a loss occurs whilst measuring
the old photons in the computational basis then measurement of any one of the
remaining old physical qubits (or indeed one from each of the new pair of
encoded parity qubits) will disentangle the other parity qubit which can then
be re-encoded.

The probability that a parity qubit will be successfully encoded, without
photon loss, is given by:
\begin{equation}
P_{Qs} = \sum_{i=1}^{n-1}(\frac{1}{2} \eta_1 \eta_2)^i \eta_1^{n-i}
\end{equation}
where the size of the original parity qubit is $n$ and the probability of
detecting an old photon is given by $\eta_1= \eta_d \eta_s \eta_m$, for a
detector efficiency of $\eta_d$, a photon source efficiency of $\eta_s$ and a
memory efficiency of $\eta_m$. The probability of successfully detecting a new
photon is given by $\eta_2 =\eta_d \eta_s$. The photon source efficiency
appears in the detection efficiency of an old photon because a photon may have
been missing from the resource state used in the previous encoding sequence.
In `reading' these probabilities it pays to keep in mind that the fusion
process will succeed or fail with probability $\eta_1\eta_2/2$ and detect a
photon loss with probability $1-\eta_1\eta_2$.

Now let us consider the case of complete (unrecoverable) failure. This will
occur if there is a sequence of fusion failures and photon loss events which
result in all of the parity qubit component photons being lost without a
successful disentangling operation being carried out. The probability of this
occurring is given by:
\begin{multline}
P_{ff} = \sum_{j=1}^{n-1}(\frac{1}{2} \eta_1\eta_2)^{j-1} (1-\eta_1\eta_2)(1-\eta_1)^{n-j} \\
+ R\sum_{j=0}^{n-2}(\frac{1}{2} \eta_1\eta_2)^{j+1}\sum_{k=0}^{n-2-j}\eta_1^{k} (1-\eta_1)^{n-1-j-k} \\
+(\frac{1}{2} \eta_1 \eta_2)^{n-1}(1-\eta_1) 
\end{multline}
Where $R=\sum_{k=1}^{q}{q \choose k}(1-\eta_2)^{kn}[1-(1-\eta_2)^{n}]^{q-k}$
and takes into account failure to decouple using the new parity qubits also
(measuring the components in diagonal basis).  That leaves the 
probability that a photon loss occurs in the encoding of one parity qubit but
that we successfully disentangle it from the other parity qubit in the
redundancy code: $P_{Qf} = 1-P_{Qs}-P_{ff}$.

We can now calculate the threshold for the memory circuit. There are two ways
in which the circuit can succeed. First, one of the parity qubits can be
encoded without photon loss and then successfully disentangled from the other.
This will occur with probability $P_{Qs}[1-(1-\eta_1)^n]$. Secondly, 
 a parity qubit can suffer photon loss but be successfully disentangled,
where-upon another parity qubit is successfully re-encoded. This will occur
with probability $P_{Qf} P_{Qs}$. Thus the probability of one successful
sequence of the memory circuit for $q$ parity qubits is:
\begin{equation}
P_{E} = \sum_{j=0}^{q-1} P_{Qf}^{j} P_{Qs}[1-(1-\eta_1)^n]^{q-1-j}  
\end{equation}
Although for fixed $n$,
$\lim_{q\rightarrow\infty}P_{E}=0$ and for fixed $q$,
$\lim_{n\rightarrow\infty}P_{E}=0$ numerical investigations indicate that it's
still possible to find $n$ and $q$ so that $P_{E}$ approaches one. 

\begin{figure}[htb]
\begin{center}
\includegraphics[width=.4\textwidth]{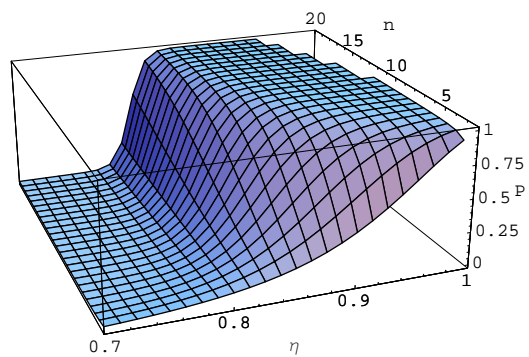}
\caption{$P_{E}$ for optimal $q$. }
\label{fig:limits}
\end{center}
\end{figure}

The optimal $q$ can be found from $\frac{d}{dq}P_{E}=0$ and using this value
numerically 
we find that $P_{E}$ approaches one for increasing $n$ provided the threshold $\eta>0.82$ is satisfied.
For efficiencies above about $0.96$ a polynomial overhead in the code size results in an exponential decrease in the failure probability ($1-P_{E}$). For lower efficiencies the overhead is exponential. In figure~\ref{fig:limits} we show the behaviour of $P_{E}$ for optimal $q$
as a function of $\eta$ and $n$. 

\textbf{Resources:} 
We now discuss the creation of the resource states used to implement our memory
circuit and hence the overheads needed. To this end we introduce a second
operation, the {\it single rail} partial Bell measurement \cite{KLM01}. This is
achieved by mixing one of the polarization modes from each of 2 physical qubits
on a beamsplitter and counting photons at the outputs. A successful event
occurs when one and only one photon is counted, otherwise it is unsuccessful.
When successful it projects onto single -rail Bell states in which a logical
zero is represented by the vacuum and a logical one by a single photon state.
In terms of dual rail qubits its effect is to project onto the states
$(\ket{H}\bra{HH}+\ket{V}\bra{VV})/\sqrt{2}$ or
$(\ket{H}\bra{HH}-\ket{V}\bra{VV})/\sqrt{2}$ when successful, and measures each
qubit in the computational basis when it fails. We will refer to this operation
as type-I fusion, ($f_I$) \cite{Bro04}.

We will take as our basic resource the Bell state $\Ket{0}{2}$.
Non-deterministic sources for such states are currently available and
considerable effort is being made to create deterministic, or at least heralded
sources of these states. To create the state $\Ket{0}{3}$, two $\Ket{0}{2}$ can
be fused
together using the $f_I$ gate. When successful, the $\Ket{0}{3}$
state is produced, when unsuccessful, both Bell states are
destroyed. Since $f_I$ functions with a probability of $1/2$, on
average two attempts are necessary, so on average each $\Ket{0}{3}$
consumes $4 \Ket{0}{2}$. 

Once there is a supply of $\Ket{0}{3}$ states, either $f_I$ or
$f_{II}$ can be used to further build up the resource state via 
\begin{align}
	(H\!\otimes\! H) f_I H \Ket{0}{n}\Ket{0}{m}&\rightarrow \left\{ \begin{array}{cl}
		\Ket{0}{m+n-1} & \mbox{(success)}\\
		- & \mbox{(failure)}
	\end{array}\right.\label{eq:fIjoin}	
\end{align}
and Eq.~\ref{encoding}. Using $f_I$ with Hadamard gates has the advantage of
losing only a single qubit from the input states, but the disadvantage of
completely destroying the encoding in both input states in the event of
failure. Using $f_{II}$ to join the input states is at the expense of losing
two of the initial qubits. There are two advantages to using $f_{II}$ ---
firstly, in the case of failure, we do not destroy the encoding so-far
produced, just reduce this encoding by one and; secondly, the operation is
``fail-safe'' in that a detection loss event is immediately recognizable as a
failure (as 2 photons will not be counted) in contrast to $f_I$ where photon
loss can lead to a false positive.

We can avoid the problem of the $f_I$ failure mode in the following way. If
$f_I$ gives a false positive it means that the mode exiting the fusion gate
does not contain a photon. Thus our supply of  $\Ket{0}{3}$ states each have
one ``suspect'' mode which may be vacuum. We now simply fuse two $\Ket{0}{3}$
with $f_I$ to form a $\Ket{0}{5}$ using the suspect modes as the fusion point.
We now are able to produce a supply of $\Ket{0}{5}$ states which again have one
suspect mode each. Finally we use $f_{II}$ to fuse two $\Ket{0}{5}$ to produce
a $\Ket{0}{8}$, once again using the suspect modes as the fusion point. This
final fusion can not give a positive if a photon had been lost in either of the
previous fusion events. In this way we can reliably produce the
resource state, $\Ket{0}{8}$, regardless of detection efficiency. Of course
missing photons due to finite source efficiency can still occur and are
accounted for by $\eta_s$ .

Using this approach and recycling $f_{II}$ failures carries an average cost of
approximately $44 \Ket{0}{2}$ per $\Ket{0}{8}$, where we have assumed high
detection and source efficiencies. Producing the encoder resource requires
first the production of a $\Ket{0}{3}$ onto which two $\Ket{0}{8}$ are fused
using $f_{II}$. A simple recycling strategy leads to a cost of approximately
$169 \Ket{0}{2}$. This is not necessarily optimal. Increasing the redundancy in
the encoder resource requires only a linear overhead, i.e. the resource state
for a q-fold redundancy encoder costs approximately $(q-1)\times 169
\Ket{0}{2}$. Increasing $n$ similarly carries a linear overhead.
%

\textbf{Conclusion}
In this paper we have introduced optical qubits with fault tolerance to loss
under linear optical manipulations.  We numerically determine the threshold for
an optical memory based on these qubits to be $82\%$ efficiency. That is, in
principle, for efficiencies higher than this threshold, it is possible to find
a suitable encoding such that the probability of a successful sequence of the
quantum memory is arbitrarily close to $1$. If we restrict ourselves to two
parity qubits each encoded across five physical qubits, and ask only when our
quantum memory works with higher probability than a passive memory, then the
answer is that the efficiencies of the sources and detectors must exceed
$96\%$.

The parity encoding we use was first introduced by KLM, however by using
incremental encoding techniques and the fusion technique we dramatically reduce
the resource usage and increase the threshold over the original scheme.
Although we have only specifically discussed a quantum memory the techniques
can be generalized to include gate operations. We expect a number of the
techniques described here could also be useful in optical quantum information
processing with non-linearities and other quantum information platforms.


We would like to acknowledge helpful discussions with Bill Munro and 
Stefan Scheel.


\end{document}